\documentclass[eqsecnum,aps,superscriptaddress,11pt ]{revtex4}
\usepackage{graphicx}
\usepackage{dcolumn}
\usepackage{xspace}
\usepackage{rotating}

\begin{document}
\title{Theoretical spectroscopic studies of the atomic transitions and lifetimes of low-lying states in Ti IV }
\vspace*{0.5cm}

\author{Subhasish Mandal$^1$, Gopal Dixit$^1$, B.K. Sahoo$^2$, R.K. Chaudhuri$^3$, and Sonjoy Majumder$^1$ \\
\vspace{0.3cm}
{\it $^1$Department of Physics, Indian Institute of Technology-Madras, Chennai-600036, India} \\
$^2$ Max Planck Institute for the Physics of Complex Systems, N\"othnitzer stra{\ss}e 38, D-01187 Dresden,
Germany\\
$^3$ Indian Institute of Astrophysics, Bangalore-34, India}
\date{\today}

\begin{abstract}
\noindent

The astrophysically important electric quadrupole (E2) and magnetic dipole (M1)
transitions for the low-lying states of triply ionized titanium (Ti IV) are
calculated very accurately using a state-of-art
all-order many-body theory called Coupled Cluster (CC) theory in the relativistic frame-work.
Different many-body correlations of the CC theory has been estimated by studying the core and
valence electron excitations to the unoccupied states. The calculated excitation energies of different
states are in very good agreement with the measurements.
Also we compare our calculated electric dipole (E1) transition amplitudes of few transitions with
recent many-body calculations by different groups. We have also carried out the calculations for the lifetimes of 
the low-lying states of Ti IV. A long lifetime is found for the first excited 3d$^{2}D_{5/2}$ state, which 
suggested that Ti IV may be one of the useful candidates for many important studies. 
Most of the results reported here are not available in the literature, to the best of our knowledge. 

\end{abstract}
\maketitle

\section{Introduction}

With the advent of improved technologies in both observational instruments in astronomy and laboratory instruments
in plasma and atomic research, accurate transitions rates ve been  possible in high resolution spectrums.
However, in many of the cases, especially for forbidden transitions of stripped electronic systems experimental
measurements are difficult. Whereas, those transitions are very important in various fields of science and technology.
Therefore, there are urgent requirement for accurate theoretical estimations  for these transitions
rates to meet the demands. 
The forbidden lines provide important clues in other areas of astrophysics, beacuse of the long lifetime of the upper
state against radiative decay. These lines are particularly sensitive to the collisional de-excitation and serve as indicators
of electron density and temparature, $N_e$ and $T_e$, in the emmision region. Determination of $N_e$ and $T_e$ from the forbidden line
intensities was discussed origanally for the general case by Seaton \cite{seaton} and Seaton and Osterbrock \cite{seaton1}.
A number of such transitions have been observed in the ultraviolet
spectrum of the solar corona. Forbidden atomic emmision lines are commonly observed in quasars with an intensity often
comparable to accompanying `allowed transition' \cite{burbidge}. Moreover, gaseous nebulae exhibit in their spectra forbidden transition lines of low excitation energy.
Many astrophysical phenomena like coronal heating, evolution of chemical composition in stellar envelopes, determination of the
chemistry in the planetary nebulae precursor's envelope are believed to be explained largely by these forbidden lines.
In laboratory tokamak plasmas and in various astronomical objects, suitably chosen these forbidden lines serve as a basis
for reliable electron density and temperature diagnostics \cite{biemont}.
Titanium is observed in a variety of stellar objects, like in  the Sun where Ti figures in third place in terms
of number of lines \cite{jaschek}. Various ionization stages  of Ti  are present in stellar plasma, for instance
in the $\tau$ Sco spectrum \cite{rogerson}. Recently, the lines of triply ionized titanium (Ti IV)  have been 
detected in Wolf-Rayet Star \cite{wolf-Rayet}. Also, Ti IV in oxidized form used in dark and photo induced decomposition of ozone in air
has been studied \cite{J.Chem}. Accurate estimation of forbidden transitions of this ionized system are
urgent to explain and quantify the band structure of the energy level of this system doped in crystal
material. Doping of Ti IV in crystal material is used to build optical and polymer devices
\cite{RCI}.

In this work, we employ the multi-reference (MR) Fock-space Coupled Cluster
(FSCC) method with single (S), double (D) and partial triple (T) excitation
to compute forbidden transitions in Ti IV using relativistic orbitals/
spinors.  The coupled-cluster (CC) theory is non-perturbative in nature and its
relativistic extension has been successfully employed earlier in many
sophisticated problems \cite{bijaya1,bijaya2} to estimate various tiny effects
precisely. In the present work, we investigate the core and valence electrons
correlation contributions obtained from the MR-FSCCSD(T) method to the M1 and
E2 transitions among the low-lying states in the considered system.

Because of its complexity and computational limitations, we have considered only the
single, double and partially triple excitations in the CC theory. However, due to its all-orderness and also because the
contributions from higher order excitations diminishes gradually, this is a quite accurate calculation.
Some of the detail applications in astrophysics and scattering physics of Ti IV
are discussed in recent paper by Kingston and Hibbert \cite{hibbert}. They have used non-relativistic
configuration interaction method for electromagnetically allowed transitions and considered relativistic effect
with Breit-Pauli approximation. 
Length and velocity forms of the transition amplitudes are compared
for few allowed transitions to check the accuracy of the numerical
approaches used.

\section{Theory}
{\bf Coupled Cluster (CC) theory for one electron attachment process}\\

The CC theory in the relativistic framework can be extended open-shell theory based 
on the no-virtual-pair approximation (NVPA) along with appropriate 
modification of orbital form and potential terms \cite{eliav}. Relativistic
CC theory begins with Dirac-Coulomb Hamiltonian (H) for an $N$ electron 
atom.  The Fock-space multireference CC (FSMRCC) approach used for any valence orbitals is
employed here. 

Since the FSMRCC theory has been described elsewhere \cite{lindgren,mukherjee,haque,Pal},
we provide a brief review of this method. The FSMRCC theory for single valence orbital
is based on the 
concept of the common vacuum for both the N and N+1 electron systems, which 
allows us to formulate a direct method of excitation energies. In this method
the holes and particles are defined with respect to the common vacuum for 
both the N and N+1 electron systems. The model space of a (m,n) Fock-space 
contains determinants with $m$ holes and $n$ particles distributed within 
a set of what are termed as {\em active} orbitals. For example, in this 
present paper, we are dealing with a (0,1) Fock space, which is a complete 
model space (CMS) by construction and is given by

\begin{equation}\label{eq4}
|\Psi^{(0,1)}_\mu\rangle=\sum_i {\mbox C}_{i\mu} |\Phi_i^{(0,1)}\rangle,
\end{equation}
\noindent
where ${\mbox C}_{i\mu}$'s are the coefficients of $\Psi^{(0,1)}_\mu$
and $\Phi^{(0,1)}_i$'s are the model space configurations.
The dynamical electron correlation effects are introduced through the
{\em valence-universal} wave-operator $\Omega$
\cite{lindgren78,mukherjee} 

\begin{equation}\label{eq5}
\Omega={\{ \exp({\tilde{S}}) }\},
\end{equation}
\noindent
where
\begin{equation}\label{eq6}
{\tilde{S}}=\sum_{k=0}^m\sum_{l=0}^n{S}^{(k,l)}
                 ={S}^{(0,0)}+{S}^{(0,1)}+ {S}^{(1,0)}+\cdots.
\end{equation}
Here $\{\cdots\}$ stands for the normal odering of the creation and annihilation operators 
related with corresponding excitations operators, $S$.
For example, the normal ordered form of the Dirac-Coulomb Hamiltonian used here is given by
\noindent 
\begin{equation}\label{eq2}
{\mathcal H}={\mbox H}-\langle\Phi|{\mbox H}|\Phi\rangle
           ={\mbox H}-{\mbox E}_{\tiny\mbox{DF}}=
\sum_{ij}\langle i|f|j\rangle\left\{ a_{i}^{\dagger}a_{j}\right\} 
+\frac{1}{4}\sum_{i,j,k,l}\langle ij||kl\rangle
\left\{ a_{i}^{\dagger}a_{j}^{\dagger}a_{l}a_{k}\right\},
\end{equation}
\noindent 
where
\begin{equation}\label{eq3}
\langle ij||kl\rangle=\langle ij|\frac{1}{r_{12}}|kl\rangle
-\langle ij|\frac{1}{r_{12}}|lk\rangle.
\end{equation}
Here E$_{\tiny\mbox{DF}}$ is the Dirac-Fock energy, $f$ is the one-electron 
Fock operator, $a_i$ $ (a_i^{\dagger})$ is the annihilation (creation) operator 
(with respect to the Dirac-Fock state as the vacuum) for the $i$th electron. 

At this juncture, it is convenient to single out the core-cluster
amplitudes $S^{(0,0)}$ and call them $T$. The rest of the
cluster amplitudes will henceforth be called $S$. Since $\Omega$
corresponding to the valence orbital $v$ is in normal order, we can rewrite Eq.(\ref{eq5}) as
\begin{equation}\label{eq7}
\Omega ={exp(T)}{\{\mbox{exp}({S_v}) }\}.
\end{equation}
Now, if we define 
\begin{equation}\label{eq9}
{\mbox H}_{\tiny\mbox {eff}}={\mbox P}^{\mbox{(k,l)}}{\mbox H}
 \Omega_v {\mbox P}^{\mbox{(k,l)}}.
\end{equation}
with the operator $P^{\mbox{(k,l)}}$ is the model space projector for k-hole and l-particle,
which satisfying complete model space condition.
The ``valence-universal'' wave-operator $\Omega$ in Eq.(\ref{eq7}) is
parametrized in such a way that the states generated by its action on the 
reference space satisfy the Fock-space Bloch equation
\begin{equation}\label{eq8}
{\mbox H}\Omega {\mbox P}^{\mbox{(k,l)}}=\Omega {\mbox P}^{\mbox{(k,l)}}.
                    {\mbox H}_{\tiny\mbox {eff}} {\mbox P}^{\mbox{(k,l)}}
\end{equation}
\noindent
To formulate the theory for direct energy differences, we pre-multiply 
Eq.(\ref{eq8}) by e$^{{-T}}$  and get
\begin{equation}\label{eq11}
{\overline{\mbox H}}\Omega_v {\mbox P}^{\mbox{(k,l)}}
      =\Omega_v {\mbox P}^{\mbox{(k,l)}}
                    {\mbox H}_{\tiny\mbox {eff}} {\mbox P}^{\mbox{(k,l)}}
\hspace{0.2in}\forall  (k,l)\ne(0,0),
\end{equation}
where ${\overline{\mbox H}}$=e$^{{-T}}$ H e$^{T }$. Since 
${\overline{\mbox H}}$ can be partitioned into a connected operator 
${\tilde{\mbox H}}$ and E$_{\tiny\mbox{ref/gr}}$ (N-electron closed-shell 
reference or ground state energy), we likewise define 
${\tilde{\mbox H}}_{\tiny\mbox{eff}}$ as
\begin{equation}\label{eq12}
{\mbox H}_{\tiny\mbox {eff}}={\tilde{\mbox H}}_{\tiny\mbox{eff}} + 
{\mbox E}_{\tiny\mbox{ref/gr}}.
\end{equation}
Substituting Eq. (\ref{eq12}) in Eq. (\ref{eq11}) we obtain the Fock-space Bloch
equation for energy differences:
\begin{equation}\label{eq13}
{\tilde{\mbox H}}\Omega_v {\mbox P}^{\mbox{(k,l)}}
      =\Omega_v {\mbox P}^{\mbox{(k,l)}}
       {\tilde{\mbox H}}_{\tiny\mbox {eff}}{\mbox P}^{\mbox{(k,l)}}.
\end{equation}
Eqs. (\ref{eq8}) and (\ref{eq13}) are solved by Bloch projection method, 
involving the left projection of the equation with P$^{\mbox{(k,l)}}$ and 
its orthogonal complement Q$^{\mbox{(k,l)}}$ to obtain the effective 
Hamiltonian and the cluster amplitudes, respectively. 

In this article,
triple excitations are included in the open shell CC amplitude which
correspond to the correlation to the valence orbitals, by an approximation
that is similar in spirit to CCSD(T) \cite{ccsd(t)}. The approximate
valence triple excitation amplitude is given by

\begin{equation}
{S^{(0,1)}}_{abk}^{pqr}=\frac{{\{{\overbrace{V{T}_2}}\}_{abk}^{pqr}}+{\{{\overbrace{V{S^{(0,1)}}_2}}}\}_{abk}^{pqr}}{\varepsilon_{a}+\varepsilon_{b}+\varepsilon_{k}-\varepsilon_{p}-\varepsilon_{q}-\varepsilon_{r}},\label{eq21}\end{equation}
where ${S^{(0,1)}}_{abk}^{pqr}$ are the amplitudes corresponding to the simultaneous
excitation of orbitals $a,b,k$ to $p,q,r$, respectively;
$\overbrace{V{T}_2}$ and $\overbrace{V{\mbox S^{(0,1)}}_2}$ are the connected
composites involving $V$ and $T$, and $V$ and $S^{(0,1)}$, respectively, where
$V$ is the two electron Coulomb integral and $\varepsilon$'s are the
orbital energies.

\section{Computational procedure}

The transition matrix element due to any operator $D$ is evaluated in 
the CC method by expressing it as
\begin{eqnarray}
D_{fi}& = & \frac{\langle \Psi_f|D|\Psi_i\rangle} 
{\sqrt{{\langle \Psi_f|\Psi_f\rangle}{\langle \Psi_i|\Psi_i\rangle}}} 
\nonumber \\
&=& \frac{{\langle \Phi_f|\{1+{S^{(0,1)}_f}^{\dag}\}{e^T}^{\dag}De^T\{1+S^{(0,1)}_i\}|\Phi_i\rangle}} 
{\sqrt{{\langle \Phi_f|\{1+{S^{(0,1)}_f}^{\dag}\}{e^T}^{\dag}e^T\{1+S^{(0,1)}_f\}|\Phi_f\rangle}
{\langle \Phi_i|\{1+{S^{(0,1)}_i}^{\dag}\}{e^T}^{\dag}e^T\{1+S^{(0,1)}_i\}|\Phi_i\rangle}}}.
\end{eqnarray}

Here, only consideraion comes from single power of the $S^{(0,1)}$ operator 
with $S^{(0,1)}_1$ and $S^{(0,1)}_2$
representing single excitation operators from valence orbital and double
excitations from core-valence orbitals, respectively.
Interesting correlation features of the transition operator $D$ are found 
in the contraction of $\overline{D}$ with $S^{(0,1)}_1$ and $S^{(0,1)}_2$, 
which represent single excitation operators from valence orbital and double 
excitations from core-valence orbitals, respectively. Since the considered 
system is a single valence system, only one power of the $S^{(0,1)}$ operator 
will contribute in this CCSD(T) calculation.  

For computational simplicity, we express $\overline{D}$ as effective terms
using the generalized Wick's theorem \cite{lindgren} as
\begin{eqnarray}
\overline{D} &=& (e^{T^{\dagger}} D e^T)_{f.c.} + (e^{T^{\dagger}} D e^T)_{o.b.} +
(e^{T^{\dagger}} D e^T)_{t.b.} + ....,
\end{eqnarray}
where we have used the abbreviations $f.c.$, $o.b.$ and $t.b.$ for fully
contracted, effective
one-body and effective two-body terms respectively. In this expansion of
$\overline{D}$,
the effective one-body and two-body terms are computed keeping terms of the
form of
\begin{eqnarray}
\overline{D}_{o.b.} &=& D + T^{\dagger} D + D T + T^{\dagger} D T ,
\end{eqnarray}
and
\begin{eqnarray}
\overline{D}_{t.b.} = D T_1 + T_1^{\dagger} D + D T_2 + T_2^{\dagger} D,
\end{eqnarray}
respectively. Other effective terms correspond to higher
orders in the residual Coulomb interaction and hence they are neglected in the
present calculation.

The reduced matrix element corresponding to E1, E2 and M1 transitions are 
given earlier papers written by few of the authors \cite{sahoo04,sahoo06}.
The emission transition probabilities (in $sec^{(-1)}$) for the E1, E2 and M1 channels from states {\it f} to {\it i} are given by
\begin{equation}
A^{E1}_{f\rightarrow i} = \frac{2.0261\times10^{18}}{\lambda^{3}[j_f]}S^{E1}_{f\rightarrow i}
\end{equation}
\begin{equation}
A^{E2}_{f\rightarrow i} = \frac{1.11995\times10^{18}}{\lambda^{5}[j_f]}S^{E2}
\end{equation}
\begin{equation} 
A^{M1}_{f\rightarrow i} = \frac{2.69735\times10^{13}}{\lambda^{3}[j_f]}S^{M1} ,
\end{equation}
where $[j_f] = 2j_f+1$ is the degeneracy of a $ f$-state, $S$ is the square of the transition 
matrix elements of any of the corresponding transition operator $D$, and $\lambda$ (in \AA) are the 
corresponding transition wavelength.

\section{RESULT AND DISCUSSIONS}

Many-body calculations started with closed shell coupled cluster calculations of Ti V. The reference state of this closed
shell system is obtained from Dirac-Fock (DF) calculation using Gaussian type orbital (GTO) formalism \cite{9}.
The exponent of the GTO functions are obtained from universal even temporing condition  with
$\alpha=0.00825$ and $\beta=2.73$ for all the symmetries. The number basis function used in this DF calculations are
32, 30, 25, 20, 20 for l=0, 1, 2, 3, 4 symmetries. Number of DF orbitals corresponding to these symmetries used in the closed
shell CC calculations are 11, 9, 8, 8 and 6. Number of active orbitals for different symmetries used in this calculations
are based on convergent criteria of core correlation energy for which it satisfies numerical completeness.

\begin{table}[h]
\caption{ Ionisation Potential(IP) in $cm^{-1}$ of different levels  of Ti IV  and its comparison with NIST value and MCHF values and the Fine Structure Splitting(FSS)}
  \begin{tabular}{lrrrrrr}
\hline
  & \multicolumn{3}{c}{IP} & \multicolumn{3}{c}{FSS} \\
\hline
States       & NIST        & MCHF   & CC  &  NIST & MCHF  & CC \\
\hline
$3d_{3/2}$  & 0 &0  &0  &  &  &   \\
$3d_{5/2}$   & 382.10   & 790.11  &418.02   &382.10   &790.11  & 418.02  \\
$4s_{1/2}$        &80388.92       &       &79716.67 &  &  &\\
$4p_{1/2}$& 127921.36   & 124749.38  & 127689.51 &  &  &\\
$4p_{3/2}$ & 128739.59  & 125539.49  & 128534.43 &818.23   &790.10  &844.91\\
$4d_{5/2}$        &196889.96   &  & 197050.03 & 85.69 &  & 96.80\\
$5s_{1/2}$           &212407.34       &        & 212823.15 &  &  &\\
$5p_{1/2}$ & 230608.89 & 228714.51  &231061.48   &315.49 & 263.37 & 353.46 \\
$5p_{3/2}$        &230924.38       &228977.88      &231414.94& & &\\
$4f_{5/2}$        &236135.29       & 234881.75       & 236217.07  &  & & \\
$4f_{7/2}$   &236142.30  &235254.86     &236220.0 &7.01 & 373.10 & 3.71\\
$5d_{3/2}$         &258838.48        &       &260290.41 &  &  & \\
$5d_{5/2}$      &258877.08       &       & 260335.26 &38.94 & & 44.85 \\
$ 6s_{1/2}$        &265847.42       &       &267187.94 &  &  &\\
$6p_{1/2}$         &274726.29        &272719.18       & 275396.29 &  &  &\\
$6p_{3/2}$   &274881.21  &272828.92  &275620.04 &154.92 & 109.74 & 223.75 \\
$5f_{5/2}$         &275847.01    & 276669.72      &277647.10 & & &\\
$5f_{7/2}$   &275861.94 &277942.67 &277633.09 &14.93 & 1272.95 &14.01 \\
$5g_{7/2}$    &278510.63        &        &278530.47 &  &  &\\
$5g_{9/2}$          &278511.23        &        &278531.05 & 0.60  &  &0.58\\
$6d_{3/2}$       & 289185.99       &       &292720.83 &  &  &\\
$6d_{5/2}$       & 289206.93      &        &292760.30  & 20.94 &  &60.30\\
\hline
\label{tab:front3}
\end{tabular}
\end{table}

In Table I, we have shown the ionisation potential obtained using the CCSD(T) method of  a few low-lying excited states
taking $3d_{3/2}$ as a ground state.  Kingston and Hibbert \cite{hibbert} have also calculated few of them
by  multiconfiguration Hartree-Fock (MCHF) method. Our calculated results are in better agreement with the experiemnetal
results (obtained from NIST \cite{nist}) in comparision with the MCHF results.
Except for $3d_{5/2}$  state, the average deviation  with the NIST results is only 0.427\%, whereas in  the MCHF method
it is 1.08\%. The CC calculated fine structure splitting (FS) of $3d$ has far better agreement than  MCHF calculation.
Also, the excellent agreement of the FS splittings of $F$ states indicates the accurate description of correlation in
the CC approach.  Especially, the all order considerations of core-polarization and pair-correlations.

Large lifetime has been estimated for $3d_{5/2}$ state as seen in table II shows its potentiality as a candidate for 
plasma temperature disgonistics in stars and plasma fusion devices. The millisecond lived excited state $4s$ might have
importance in  many astronomical diagonistics. \\
\begin{table}[h]
\caption{The lifetime(in $Sec$) of few low-lying states}
\begin{tabular}{lrr}
\hline
States & Lifetime\\
\hline
$3d_{5/2}$ &1.274E+03\\
$4s_{1/2}$ &7.531E-04 \\
$4p_{1/2}$ &4.651E-10\\
$4p_{3/2}$ &4.563E-10\\
\hline
\end{tabular}
\end{table}

Table III provides the comparison of the CC calculated electric dipole (E1) oscillator strengths (f-value) with the MCHF 
\cite{hibbert} in length and velocity form. In most of the cases 
MCHF underestimate the  f-values, though there are cases where good agreement seen among the results obtained from both
the methods. The good agreement  between the results of length and velocity forms indicates the accuracy of the
numerical approaches employed. 

\begin{table}[h]
\caption{ Oscillator strengths in the length $f_{l}$ and velocity $f_{v}$ form for  $ E_{1}$  transitions and its comparison with MCHF results\cite{hibbert}.}
\begin{tabular}{llrrrr}
\hline
 Transitions     &                       & $f_{l}$(MCHF) & $f_{v}$(MCHF)  & $f_{l}$(CC) & $f_{v}$(CC)\\
\hline
$3d_{3/2}$&$\rightarrow 4p_{1/2}$ & 0.0765 &0.0914&0.1588   & 0.1103\\
          &$\rightarrow 4p_{3/2}$ & 0.0154 &0.0182      &0.0158 & 0.0109\\
          &$\rightarrow 5p_{1/2}$ & 0.0080 & 0.0091       &0.0185 &0.0129 \\
          &$\rightarrow 5p_{3/2}$ & 0.0016 & 0.0019     & 0.0037 & 0.0012\\
          &$\rightarrow 6p_{1/2}$ &0.0030 &0.0031       &0.0075&0.0042\\
          &$\rightarrow 6p_{3/2}$ & 0.0006 &0.0007       &0.0022&0.0014\\
          &$\rightarrow 4f_{5/2}$ &0.1248 & 0.1109       & 0.1020&0.1066\\
$3d_{5/2}$&$\rightarrow 4p_{3/2}$  &0.0925   &0.1093 &0.1430&0.0982\\
          &$\rightarrow 5p_{3/2}$ & 0.0011 &0.0111       &0.0017&0.0112\\
          &$\rightarrow 6p_{3/2}$ &  0.0038 &0.0039       &0.0076&0.0040\\
          &$\rightarrow 4f_{5/2}$ &  0.0060 &0.0053       &0.0049&0.0041\\
          &$\rightarrow 4f_{7/2}$ & 0.1200 &0.1069     &0.1108&0.1154\\
\hline
\label{tab:front3}
\end{tabular}
\end{table}

Table IV  presents the electric quadrupole and magnetic dipole  
transition wavelengths and amplitudes, respectively, for most of the
low-lying states. They are all relevant to astrophysically studies.
The calculated wavelengths have good agreement for most of the cases
with the result obtained from the website of National Institute of
Standard and Technology (NIST) \cite{nist}. From physics point of view,
the important transitions among these are the forbidden transitions
among the fine-structures of the $3d$ and $4p$ states. Former one
falls in the infrared region, which has many applications in the plasma
research and infrared laser spectroscopy \cite{thogersen}. The latter one
falls in the optical region, has immense prospect in different atomic
physics experiments.  We have not reported wavelengths  for most of other
fine structure transitions  those fall far beyond the infrared region. \\

Quantitative contributions from different correlation terms for few  E2
transitions among low-lying states are presented in Table V.
The table shows a comparative estimations of core-polarization,
core-correlation and pair-correlation effects in these transitions. The
diagrams involving these contributions are discussed in our earlier papers
\cite{sonjoy3}.
Though all order effect
of core-polarization and pair correlation contributions are considered in the calculations here. Table shows the lowest
order contributions of them for few transitions among the low-lying states. The unusual strong core correlation, almost
same as DF, contribution has been seen for E2 transition among the fine structure states of $4p$. Core correlation
are weakest among the three correlations presented in the table.
Dominance of pair correlation effects over core polarization observed in all the transitions.

In ths similar manner the Quantitative contributions from different correlation terms for few M1 
transitions among low-lying states are presented in Table VI. From the table VI, it is really interesting to see low 
correlation effects, especially, core porization effect is almost negigible up to the digits displayed in the
table. Few cases, strong pair correlations are noticeable.

\begin{table}[h]
\caption{Transition wavelengths and transition amplitudes of Ti IV for electric quadrupole (E2)  and magnetic dipole transitions (M1)}
\begin{tabular}{llrrr}
\hline
 Transition &             & $\lambda_{CC}$ &  $ E_{2}$ & $M_{1}$ \\
\hline
$3d_{3/2}$&$\rightarrow 3d_{5/2}$ &     & -1.0336 &-1.5458 \\
          &$\rightarrow 4d_{3/2}$   & 508.81 & 1.4928  &0.0863  \\
          &$\rightarrow 4d_{5/2}$ &507.48   &  1.0186 & -0.0026\\
          &$\rightarrow 5d_{3/2}$ &384.18  &  0.5334 &0.0443\\
          &$\rightarrow 5d_{5/2}$ &384.12 &  0.3738 &-0.0012\\
            &$\rightarrow 6d_{3/2}$&341.62  &  0.3894 & 0.0371\\
           &$\rightarrow 6d_{5/2}$ &     &0.2769 &-0.0009\\
          &$\rightarrow 4s_{1/2}$ & 1254.44  & -2.1842 &\\
          &$\rightarrow 5s_{1/2}$ & 469.87  &  -0.0283 &\\
          &$\rightarrow 6s_{1/2}$  & 374.29 &  -0.0159 &\\
          &$\rightarrow 5g_{7/2}$ &359.03 & -1.1243 & \\
$3d_{5/2}$&$\rightarrow 4d_{3/2}$ &509.89   & -1.0034 & 0.0012\\
          &$\rightarrow 4d_{5/2}$ &508.56  & 1.9701 &0.0241\\
          &$\rightarrow 5d_{3/2}$ &384.80  & -0.3562 &-0.0009\\
            &$\rightarrow 6d_{3/2}$ &342.11  & -0.2606 & 0.0008 \\
          &$\rightarrow 4s_{1/2}$  & 1261.05  & -2.6593 &     \\
         &$\rightarrow 5s_{1/2}$  &470.79  & -0.0467 &   \\
          &$\rightarrow 6s_{1/2}$    & 374.85  & -0.0229 & \\
        &$\rightarrow 5g_{7/2}$  & 359.57   & 0.3775 & \\
            &$\rightarrow 5g_{9/2}$ & 359.57    & -1.3342& \\
         &$\rightarrow 6d_{5/2}$ &    &   &0.1035\\
         &$\rightarrow 5d_{5/2}$ & 384.73  &  & 0.1239\\
$4d_{3/2}$&$\rightarrow 4d_{5/2}$ &  &  -9.1095 &-1.5485\\
           &$\rightarrow 5d_{3/2}$ & 1568.49 &  7.4885 &0.1370\\
           &$\rightarrow 5d_{5/2}$ &1567.39 &  5.3203&-0.0010\\
           &$\rightarrow 6d_{3/2}$ &1039.66 & 2.3737 & -0.0010\\
          &$\rightarrow 6d_{5/2}$ & & 1.8081&-0.0004\\
          &$\rightarrow 4s_{1/2}$ &856.02  & 7.3600 &\\
          &$\rightarrow 5s_{1/2}$ &6139.37  & 14.7325 &\\
         &$\rightarrow 6s_{1/2}$  &1415.37  & -2.0579 &\\
       &$\rightarrow 5g_{7/2}$  &1219.58  &    24.1136 &\\

\hline
\end{tabular}
\end{table}

\begin{table}[h]
{\noindent (Continuation of Table IV) \hspace*{5cm} \\ }
\begin{tabular}{llrrr}
\hline
 Transition &             & $\lambda_{CC}$ &  $ E_{2}$ & $M_{1}$ \\
\hline
$4d_{5/2}$&$\rightarrow 5d_{3/2}$ & 1568.49 & -5.3540 &-0.0022\\
            &$\rightarrow 5d_{5/2}$ &1567.39 &9.8113 &0.3836\\
             &$\rightarrow 6d_{3/2}$ &1039.66 &-1.8214 &--0.0015\\
           &$\rightarrow 6d_{5/2}$  &6139.51 &3.0965 &0.2316\\
           &$\rightarrow 4s_{1/2}$ & 856.03  & 8.9919 & \\
          &$\rightarrow 5s_{1/2}$  & 6139.51 & 18.0734 &\\
          &$\rightarrow 6s_{1/2}$  & 1413.37  & -2.5365 & \\
         &$\rightarrow 5g_{7/2}$  & 1219.57   &  -8.0497  & \\
         &$\rightarrow 5g_{9/2}$  &1219.57  &  28.4611 &  \\
$5d_{5/2}$&$\rightarrow 6d_{3/2}$ & 3087.79 & -15.4604 &-0.0021\\
            &$\rightarrow   6d_{5/2}$ &    &     & 0.5057\\
             &$\rightarrow  4s_{1/2}$ & 553.65 &0.9166 & \\
              &$\rightarrow  5s_{1/2}$ & 2104.73 &-30.6035 & \\
            &$\rightarrow  6s_{1/2}$ &14592.83    &58.9756 &     \\
            &$\rightarrow  5g _{7/2}$ & 5495.95 & 24.2502 & \\
            &$\rightarrow  5g _{9/2}$ & 5495.77 & -85.7563 & \\
$5d_{3/2}$&$\rightarrow 5d_{5/2}$ &          & -33.0491 & -1.5479\\
           &$\rightarrow 6d_{3/2}$ &3083.52 &28.4053 &   0.1803\\
          &$\rightarrow 6d_{5/2}$ &         & 15.3297 & -0.0019 \\
           &$\rightarrow 4s_{1/2}$ & 533.79 &0.7484 &           \\
            &$\rightarrow 5s_{1/2}$ & 2106.73 & -24.9925 &       \\
            &$\rightarrow 6s_{1/2}$ &14497.95 &48.0724 &        \\
            &$\rightarrow 5g_{7/2}$ &5482.44 &-72.6985 &     \\
$6d_{3/2}$&$\rightarrow 4s_{1/2}$ &469.47 &0.5346 &     \\
          &$\rightarrow 5s_{1/2}$ &1251.60 &-6.2888 &     \\
         &$\rightarrow 6s_{1/2}$ &3916.52 & -60.8104 &     \\
         &$\rightarrow 5g_{7/2}$ &6627.89 & 42.6236 &     \\
$6d_{5/2}$&$\rightarrow 4s_{1/2}$ &469.38 &0.6626 &     \\
          &$\rightarrow 5s_{1/2}$ &1250.98 &-7.7575 &     \\
          &$\rightarrow 6s_{1/2}$ &3910.47 &-74.4313 &     \\
          &$\rightarrow 5g_{5/2}$ &7027.49 &-14.1738 &     \\
          &$\rightarrow 5g_{9/2}$ &7027.77 &50.1131 &     \\
$5g_{7/2}$&$\rightarrow 5g_{9/2}$ &  & -21.7443 &\\
$4p_{1/2}$&$\rightarrow 4p_{3/2}$ &   & -16.4408 &-1.1466 \\
          &$\rightarrow 5p_{1/2}$ &967.38  &   & -0.0248 \\
        &$\rightarrow 5p_{3/2}$ & 964.08 & -10.2894 & 0.0049\\ 
         &$\rightarrow 6p_{1/2}$  & 677.02 &   & -0.0134\\
           &$\rightarrow 6p_{3/2}$  &675.99 &   & 0.0030\\
           &$\rightarrow 4f_{5/2}$ &921.42 &-24.0090 &  \\
           &$\rightarrow 5f_{5/2}$   & 666.86 & 3.1636 &\\
\hline
\end{tabular}
\end{table}

\begin{table}[h]
{\noindent (Continuation of Table IV) \hspace*{5cm} \\ }
\begin{tabular}{llrrr}
\hline
 Transition &             & $\lambda_{CC}$ &  $ E_{2}$ & $M_{1}$ \\
\hline
$4p_{3/2}$&$\rightarrow 5p_{1/2}$ & 975.35  &  -5.3367    &0.0058\\
         &$\rightarrow 5p_{3/2}$ & 972.00 & -4.7702 &-0.1556 \\
          &$\rightarrow 6p_{1/2}$ &680.91  &  -2.1388    &0.0041\\
          &$\rightarrow 6p_{3/2}$ &  679.87 &-1.8725 &-0.0851\\
          &$\rightarrow 4f_{5/2}$ & 928.65 &6.3548 & \\
          &$\rightarrow 4f_{7/2}$ & 928.62 &-15.5754 & \\
          &$\rightarrow 5f_{5/2}$ & 670.63 &-0.8062 & \\
          &$\rightarrow 5f_{7/2}$ & 670.69 &2.1849 & \\
$5p_{1/2}$&$\rightarrow 5p_{3/2}$ &  282914.16 & -30.1611 & -1.1508 \\
        &$\rightarrow 6p_{1/2}$ &  2255.56 &      & 0.0306 \\
        &$\rightarrow 6p_{3/2}$ &  2244.24 & 17.5393& -1.1467 \\
        &$\rightarrow 4f_{5/2}$ &  19396.42 & -24.4364 & \\
        &$\rightarrow 5f_{5/2}$ & 2146.58 & -35.8189 & \\
$5p_{3/2}$&$\rightarrow 6p_{1/2}$ & 2273.69 & 17.8798    &-0.0089\\
         &$\rightarrow 6p_{3/2}$ & 2262.18 & 16.7129 & 0.1942\\
         &$\rightarrow 4f_{5/2}$ & 20824.11 & 12.8459 & \\
         &$\rightarrow 4f_{7/2}$ & 20811.39 & 2.1849 & \\
         &$\rightarrow 5f_{5/2}$ & 2177.12 & 19.3650 &\\
         &$\rightarrow 5f_{7/2}$ & 2122.42 &47.4574 &\\
$6p_{1/2}$&$\rightarrow 6p_{3/2}$ &  &-70.9679 & -1.1435\\
         &$\rightarrow 4f_{5/2}$ & 2552.37 &6.9155 &\\
         &$\rightarrow 5f_{5/2}$ &     & 76.6077 & \\
$6p_{3/2}$&$\rightarrow 4f_{5/2}$ &  2537.88 &-3.7780 & \\
         &$\rightarrow 4f_{7/2}$ & 2538.11 & 9.2581 & \\
         &$\rightarrow 5f_{5/2}$ & 399.86 & -41.2699 & \\
         &$\rightarrow 5f_{7/2}$ & 399.88 &100.9901 &  \\
$4f_{5/2}$&$\rightarrow 4f_{7/2}$ &     & -10.0372 & -1.8513\\
          &$\rightarrow 5f_{5/2}$ & 2413.71 &-14.7715 &  \\
         &$\rightarrow 5f_{7/2}$ &  2414.52 & -6.4105  & \\
$4f_{7/2}$&$\rightarrow 5f_{5/2}$ &2413.71  &6.4049 &\\
        &$\rightarrow 5f_{7/2}$ & 2414.52  & -17.4159 & \\
$5f_{5/2}$&$\rightarrow 5f_{7/2}$ &     & -35.759 & -1.8511 \\
$4s_{1/2}$&$\rightarrow 5s_{1/2}$ & 751.27 &      &  -0.0687\\
            &$\rightarrow 6s_{1/2}$  & 533.41 &        & -0.0386\\
$5s_{1/2}$&$\rightarrow 6s_{1/2}$ & 1839.42  &   & 0.0837\\

\hline
\end{tabular}
\end{table}

\begin{table}[h]
\caption{Explicit contributions from the MR-FSCCSD(T) calculations to the absolute magnitude of reduced E2 transitions matrix elements in a.u.}
\label{la}
  \begin{tabular}{llcccccc}
\hline
Transition & $DF  $ & $ Core-Correlation$ & $ Pair-Correlation$ & $ Core-Polarization $ & $ Norm$ & $Total$\\
\hline
$3d_{3/2}\rightarrow 3d_{5/2}$ & -1.1938&0.0010&0.0688&0.0794&0.0223&-1.0335\\
$3d_{3/2}\rightarrow 4d_{3/2}$ &1.5863&-0.0016&-0.1227&0.0198&-0.0234&1.4928\\
$3d_{3/2}\rightarrow 4d_{5/2}$ &1.0364&-0.0003&-0.0477&0.0385&-0.0160&1.0186\\
$3d_{3/2}\rightarrow 5d_{3/2}$ &0.5426&0.0025&-0.0469&0.0341&-0.0328&-2.1843\\
$3d_{3/2}\rightarrow 4s_{1/2}$ &-2.3347&0.0050&0.1626&-0.0269&0.0328&-2.1843\\
$3d_{3/2}\rightarrow 5s_{1/2}$ &-0.0597&0.0009&0.0076&0.0093&0.0004&-0.0283\\
$3d_{5/2}\rightarrow 4d_{3/2}$ & -1.0433&0.0001&0.0464&-0.2254&0.0157&-1.0003\\
$3d_{5/2}\rightarrow 4d_{5/2}$ &2.0825&-0.0001&-0.1589&0.0726&0.3562&1.9701\\
$3d_{5/2}\rightarrow 4s_{1/2}$ & -2.8689&0.0076&0.1983&-0.0125&-0.0401&-2.6593\\
$4d_{5/2}\rightarrow 5d_{3/2}$ & -5.5128 &0.0071  &  0.0922 &-0.0167&0.0511 &-5.3540\\
$4d_{5/2}\rightarrow 5d_{5/2}$ & 10.9993 &-0.0046  &  -1.0611& 0.0319 &-0.0978 &9.0113\\
$5d_{3/2}\rightarrow 4s_{1/2}$ & 0.9699 &-0.0051  &  -0.2398&-0.0268 &-0.0066 & 0.7484\\
$5d_{3/2}\rightarrow5s_{1/2}$ &-25.758&0.0538&0.5885& 0.0133&0.1695&-24.9925\\
$5d_{5/2}\rightarrow 5s_{1/2}$ & -0.0753&-0.0014&0.0072&-0.0786&0.0006&0.0467\\
$4p_{1/2}\rightarrow 4p_{3/2}$ & -8.6289&-8.5459&0.5297&0.0812&0.1789&-16.4408\\
$5p_{1/2}\rightarrow 4p_{3/2}$ &-5.2202&0.0137&-0.1654&-0.0293&0.0335&-5.3369\\
$5p_{1/2}\rightarrow 5p_{3/2}$ &-31.4436&0.1799&1.0952&0.0301&0.0888&-30.1611\\
\hline
\end{tabular}
\end{table}

\begin{table}[h]
\caption{Explicit contributions from the MR-FSCCSD(T) calculations to the absolute magnitude of reduced M1 transitions matrix elements in a.u.}
\label{la}
\begin{tabular}{llrrrrrr}
\hline
Transition &   $DF  $ & $ Core-Correlation$ & $ Pair-Correlation$ & $ Core-Polarization $ & $ Norm$ & $Total$\\
\hline
$3d_{3/2}\rightarrow 3d_{5/2}$ &-1.5489&0.0005&0.0000&0.0000&0.0333&-1.5458\\
$3d_{3/2}\rightarrow 4d_{5/2}$ &-0.0015&-0.0000&0.0097&0.0001&0.0000&-0.0026\\
$3d_{5/2}\rightarrow 4d_{3/2}$ & -0.0017&0.0001&-0.0099&-0.0002&0.0000&-0.0012\\
$3d_{5/2}\rightarrow 4d_{5/2}$ &-0.0007&0.0001&0.1857&-0.0000&-0.0038&0.2407\\
$3d_{5/2}\rightarrow 5d_{3/2}$ &-0.0001&-0.0001&-0.0061&-0.0001&-0.0000&-0.0008\\
$4d_{3/2}\rightarrow 4d_{5/2}$ &-1.5491&0.0003&0.0000&0.0000&0.0153&-1.5485\\
$5d_{3/2}\rightarrow 5d_{5/2}$ &-1.5492&0.0019&0.0001&0.0000&0.0147&-1.5479\\
$4s_{1/2}\rightarrow 5s_{1/2}$ &0.0004&-0.0006&0.0560&0.0000&0.0004&-0.0687\\
$4p_{1/2}\rightarrow 4p_{3/2}$ &-1.1545&0.0059&0.0000&0.0000&0.0125&-1.1466\\
$4p_{1/2}\rightarrow 5p_{1/2}$ &0.0000&0.0000&-0.0207&0.000&0.0002&-0.0248\\
$4p_{1/2}\rightarrow 5p_{3/2}$ &-0.0059&0.0108&-0.0060&0.0000&-0.0000&0.0049\\
\hline
\end{tabular}
\end{table}

\section{CONCLUSION}
   In this paper, we have reported the ionisation potential of a few  excited states 
of Ti IV by using the MR-FSCCSD(T) method,
which are in excillent  with the NIST results. Magnetic dipole and electric quadrupole
transition amplitudes among the bound states of the system are important for
astronomical observations and plasma researches. Here, we have reported these
results for the first time. Especially, forbidden transitions between the fine
structure $4p$ states may be considered for different atomic experiments of
fundamental physics due to its optical transition line. Long lifetime has been observed
for the first excited $D$- state and it can be used as potential metastable
state for experiments in physics. We have also
highlighted different correlation effects arising through the MR-FSCCSD(T)
method.


\begin{thebibliography}{99}
\bibitem{seaton}
Seaton M J 1954 {\it Mon. Not. R. Astron. Soc.} {\bf 114} 154.
\bibitem{seaton1}
Seaton M J and  Osterbrock D E 1957 {\it Astrophys. J.} {\bf 125} 66.
\bibitem{burbidge}
Burbidge G and  Burbidge M 1967 {\it Quasi-Stellar Objects} (San Francisco: W. H. Freeman).
\bibitem{biemont}
Biemont E and Zeippen C J 1996 {\it Comments At. Mol. Phys.} {\bf 33} 29.
\bibitem{jaschek}
Jaschek C and Jascheck M 1995 {\it The behavior of chemical elements in starts} Cambridge Univ Press.
\bibitem{rogerson}
Rogerson J  B  and  Ewell N  W  1985  {\it Astrophys. J Suppl. S.} {\bf 58} 265.
\bibitem{wolf-Rayet}
Destombes J P, Shephard Thorn E P, Redding J H and Morzadec Kerfourn M T 1975 {\it Phil. Trans. R. Soc. Lond. A} {\bf 279} 243.
\bibitem{J.Chem}
Ohtani B, Zhang S W, Nishimoto S I and Kagiya T 1992
{\it J. Chem. Soc. Faraday Trans.} {\bf 88} 1049.
\bibitem{RCI}
Murakami Shin-Ya, Kominami Hiroshi, Kera Yoshiya, Ikeda Shigeru, Noguchi Hidenori, Uosaki Kohei and Ohtani Bunsho
 2007 {\it Research on Chemical Intermediates}  {\bf 33} 285.
\bibitem{bijaya1}
Sahoo B K, Chaudhuri R K, Das B P and Mukherjee D 2006 {\it Phys. Rev. Letts.} {\bf
  96} 163003.

\bibitem{bijaya2}
Sahoo B K, Sur C, Beier T, Das B P, Chaudhuri R K and  Mukherjee D 2007 {\it Phys.
  Rev. A} {\bf 75} 042504.

\bibitem{hibbert}
Kingston A E and  Hibbert A 2006 {\it J. Phys. B} {\bf 39} 2217.
\bibitem{eliav}
Eliav E,  Kaldor U and  Ishikawa Y 1995 {\it Phys. Rev. A} {\bf 51} 225.
\bibitem{lindgren}
Lindgren I and  Morrison J 1985 {\it Atomic Many-body Theory} {\bf 3}, Ed. G. E.
  Lambropoulos and H. Walther (Berlin: Springer).

\bibitem{mukherjee}
Lindgren I and Mukherjee D 1987 {\it Phys. Rep.} {\bf 151} 93.

\bibitem{haque}
Haque A and  Mukherjee D 1984  {\it J. Chem. Phys.} {\bf 80} 5058.

\bibitem{Pal} 
Pal S, Rittby M, Bartlett R J, Sinha D and Mukherjee D 1987 {\it Chem. Phys. Lett.}
{\bf 137} 273;  1988 {\it J. Chem. Phys.} {\bf 88} 4357.
\bibitem{lindgren78}
Lindgren I 1978 {\it Int. J Quantum Chem. Symp.} {\bf 12} 33.
\bibitem{ccsd(t)}
Raghavachari K, Trucks G W, Pople J A and Head-Gordon M 1989  {\it Chem. Phys.
  Lett.} {\bf157} 479 ; Urban M, Noga J,  Cole S J and 
  Bartlett R J 1985 {\it Chem. Phys. Lett.} {\bf 83} 4041.
\bibitem{sahoo04}
Sahoo B K, Majumder S, Chaudhuri R K, Das B P and Mukherjee D 2004 {\it J. Phys. B} {\bf 37} 3409.

\bibitem{sahoo06}
Sahoo B K, Majumder S, Merlitz H, Chaudhuri R K, Das B P and Mukherjee D 2006, {\it J. Phys. B} {\bf 39} 355. 

\bibitem{9}
Chaudhuri R K, Panda P K, Das B P, Mahapatra U S and Mukherjee D 2000 {\it J.Phy. B} {\bf 33} 5129.
\bibitem{nist}
http://physics.nist.gov/Pubs/AtSpec/node17.html.

\bibitem{thogersen}
Thogersen J, Scheer M, Steele L D, Haugen H K and Wijesundera W P 1996 {\it Phys.
  Rev. Lett} {\bf 76} 2870.

\bibitem{sonjoy3}
Majumder S, Sahoo B K, Chaudhuri R K, Das B P and  Mukherjee D 2006 {\it Eur. Phys. J.
  D} e2006-00248-2.




\end{thebibliography}
\end{document}